\documentclass%
[prl,%
superscriptaddress,%
showpacs,%
twocolumn]
{revtex4}

\usepackage{latexsym}        
\usepackage{amssymb}
\usepackage{amsmath}
\usepackage{amsfonts}
\usepackage[dvips]{graphicx} 

\begin{document}

\preprint{APS/123-QED}

\title{Atomic-scale relaxation dynamics and aging in a metallic glass probed by X-ray photon correlation spectroscopy}
\author{B.~Ruta}%
 \email{ruta@esrf.fr}
 \affiliation{European Synchrotron Radiation Facility, BP220, F-38043 Grenoble, France.}
\author{Y.~Chushkin}%
\affiliation{European Synchrotron Radiation Facility, BP220, F-38043 Grenoble, France.}
\author{G.~Monaco}%
\affiliation{European Synchrotron Radiation Facility, BP220, F-38043 Grenoble, France.}
\author{L.~Cipelletti}
\affiliation{Universit\'{e} Montpellier 2, Laboratoire Charles Coulomb UMR 5221, F-34095, Montpellier, France.}%
\affiliation{CNRS, Laboratoire Charles Coulomb UMR 5221, F-34095, Montpellier, France.}
\author{E.~Pineda}
\affiliation{Departament de F\'{i}sica i Enginyeria Nuclear, ESAB, Universitat Polit\`{e}cnica de Catalunya, c/ Esteve Terradas 8, 08860 Castelldefels, Spain.}%
\author{P.~Bruna}
\affiliation{Departament de F\'{i}sica Aplicada, EPSC, Universitat Polit\`{e}cnica de Catalunya, c/ Esteve Terradas 5, 08860 Castelldefels, Spain.}
\author{V.~M.~Giordano}
\affiliation{LPMCN, Universit\'{e} Claude Bernard Lyon 1 and CNRS, 69622, Villeurbanne, France.}
\author{M.~Gonzalez-Silveira}%
\affiliation{%
Nanomaterials and Microsystems Group, Physics Department, Universitat Aut\`{o}noma de Barcelona, 08193 Bellaterra, Spain.}

\date{\today}

\begin{abstract}
We use X-Ray Photon Correlation Spectroscopy to investigate the structural relaxation process in a metallic glass on the atomic length scale. We report evidence for a dynamical crossover between the supercooled liquid phase and the metastable glassy state, suggesting different origins of the relaxation process across the transition. Furthermore, using different cooling rates we observe a complex hierarchy of dynamic processes characterized by distinct aging regimes. Strong analogies with the aging dynamics of soft glassy materials, such as gels and concentrated colloidal suspensions, point at stress relaxation as a universal mechanism driving the relaxation dynamics of out-of-equilibrium systems. 
\end{abstract}


\maketitle
\begin{figure}
\includegraphics[width=0.53\textwidth]{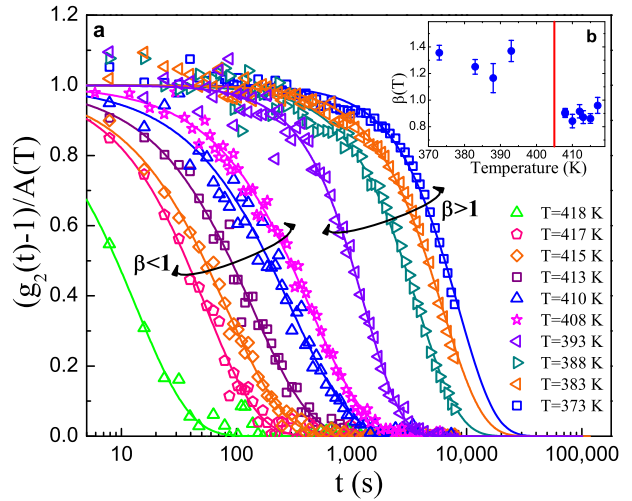}
\vspace{-0.8cm}
\caption{\label{Fig1} Temperature dependence of correlation functions measured for $q_0=2.56$ \r{A}$^{-1}$ by means of XPCS. Lines are the best fits using a KWW model function. The black arrows emphasize the dynamical crossover from stretched ($\beta<1$) to compressed ($\beta>1$) exponential functions below 408 K. The curves in the glassy state correspond to the SC-g and are measured on lowering T from the liquid phase. b, Temperature dependence of the corresponding shape parameter. The line indicates the $T_g$ obtained from calorimetric measurements (see SI).}
\vspace{-0.5cm}
\end{figure}
\begin{figure}
\includegraphics[width=0.5\textwidth]{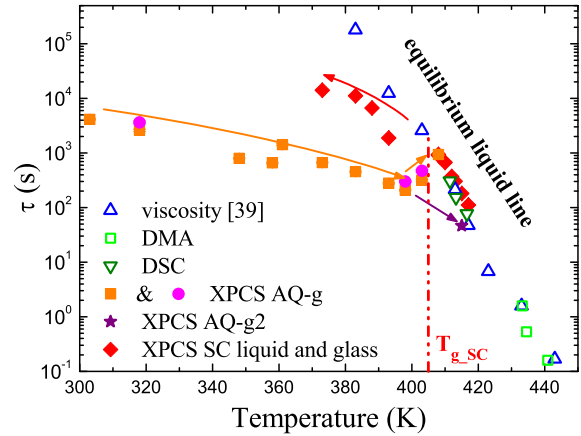}
\vspace{-0.8cm}
\caption{\label{Fig2} Structural relaxation times measured with XPCS. The dotted-dashed line indicates the $T_g$ for the SC-g. We report data for each glass at the same waiting time $t_w\sim6000$ s for the AQ-g and $t_w\sim9000$ s for the SC-g after temperature equilibration. The filled circles correspond to the AQ-g at the same $t_w\sim$9000 s as for the SC-g, for a consistent comparison.}
\vspace{-0.5cm}
\end{figure}
\begin{figure}
\includegraphics[width=0.53\textwidth]{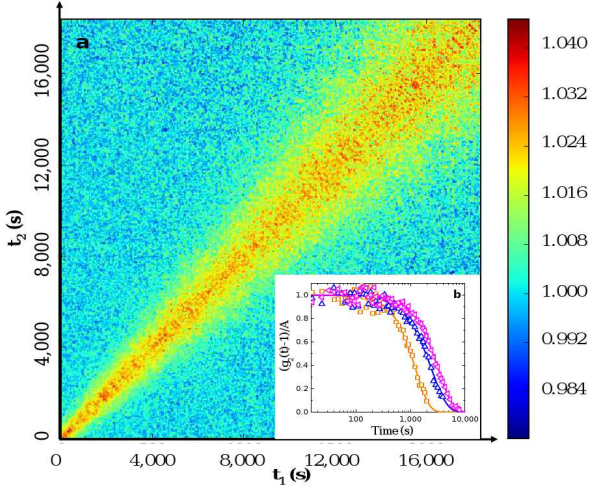}
\vspace{-0.9cm}
\caption{\label{Fig3}a, Two time correlation function measured in the AQ-g at $T$=318 K by means of XPCS. b, Corresponding $g_2(t,t_w)$ calculated for different sample ages. From left to right $t_w$=4680 s, $t_w$=9840 s and $t_w$=12960 s. The data are reported together with the best fit curves with a KWW model function.}
\vspace{-0.5cm}
\end{figure}
\begin{figure}
\includegraphics[width=0.48\textwidth]{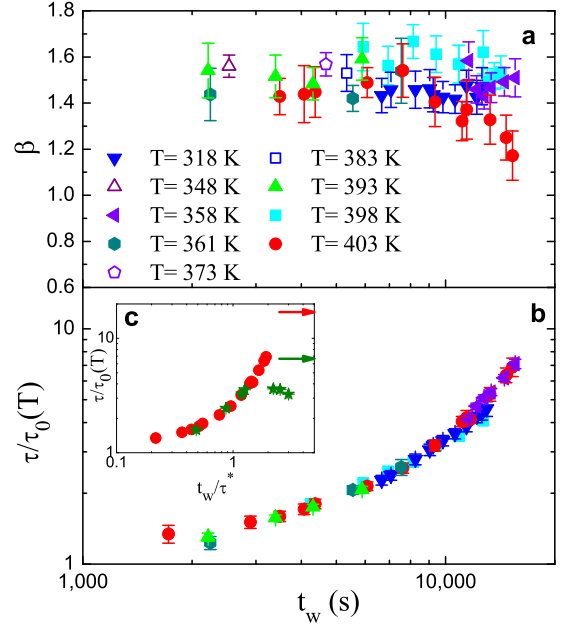}
\vspace{-0.9cm}
\caption{\label{Fig4} a, Sample age and temperature dependence of the shape parameter $\beta$ measured with XPCS in the AQ-g. b, Corresponding structural relaxation times normalized by their initial value $\tau_0$(T). Data corresponding to temperatures where only one curve has been measured are reported only in panel a (empty symbols). c, Structural relaxation times measured at $T=403$ K for both AQ-g (circles) and AQ-g2 (stars) normalized by their initial value $\tau_0$ and reported as a function of $t_w$ rescaled by the parameter $t^{\ast}$ obtained from the analysis of the fast aging regime. The data are reported together with the corresponding equilibrium values (red and green arrow, respectively) obtained by rescaling for $\tau_0$ the equilibrium value reported in Figure \ref{Fig2}.}
\vspace{-0.6cm}
\end{figure}

Glasses are usually defined as liquids that are trapped in a metastable state from which they slowly evolve toward the corresponding equilibrium phase \cite{angell1995,debenedetti2001}. Although aging is known since centuries, a clear picture of the dynamics in the glassy state is still missing \cite{angell2000b,berthier2011}. Most of the experimental information available on aging concerns macroscopic quantities, such as viscosity or elastic moduli \cite{struik1978,hodge1994,miller1997,wen2006,busch1998,hachenberg2008,khonik2009}, or focuses on dielectric relaxation \cite{leheny1998,lunkenheimer2005,casalini2009,alegria1995,alegria1997,hecksher2010}, a quantity that is often difficult to relate directly to the particle-level dynamics. From these measurements, a characteristic time for the evolution towards equilibrium can be extracted, but no direct information on the connection between aging and the underlying microscopic dynamics is available. 
By contrast, a full understanding of aging requires a detailed description of the particle-level dynamics, and in particular of the structural relaxation time $\tau$, the characteristic time for a system to rearrange its structure on the length scale of its constituents. While the structural relaxation has been widely investigated in the liquid phase, only a few studies report on the behaviour of $\tau$ below the glass transition temperature, $T_g$ \cite{casalini2009,alegria1995,alegria1997,kob1997}. Molecular dynamics simulations show that $\tau$ increases linearly or sub-linearly with the waiting time, $t_w$, and suggest the possibility of rescaling the measured quantities on a single master curve (time-waiting time superposition principle) \cite{kob1997}. Studies in this direction, however, have led to debated results \cite{struik1978,leheny1998,lunkenheimer2005,richert2010,hecksher2010}. Consequently, several key questions remain unanswered: what is the fate of the structural relaxation process when the system falls out of equilibrium in the glassy state? How does it depend on the thermal history and the waiting time? What physical mechanism is responsible for structural relaxation? \\
Here, we address these questions by presenting an experimental investigation of the structural relaxation on the atomic length scale in a metallic glass former, in both the supercooled liquid and the glassy state, as a function of temperature and waiting time, and for different thermal histories. We use X-Ray Photon Correlation Spectroscopy (XPCS) to study Mg$_{65}$Cu$_{25}$Y$_{10}$, a well-known glass former with a relatively low $T_g\sim 405$ K and a stable supercooled region.\\
The experiments were performed at the beamline ID10 of the European Synchrotron Radiation Facility. A partially coherent beam with an energy of 8 keV was selected by rollerblade slits opened to $10\times10$ $\mu m$, placed $\sim$0.18 m upstream of the sample. Mg$_{65}$Cu$_{25}$Y$_{10}$ ribbons were placed in a furnace, providing a temperature stability of 0.1 K. The scattering intensity was collected in transmission geometry by an IkonM charge coupled device from Andor Technology ($1024\times1024$ pixels, $13\times13$ $\mu m$ pixel size) placed 0.68 m from the sample. Sets of $\sim3000$ images were taken with 5 s or 7 s exposure time per frame and were analyzed following the procedure described in Ref. \cite{chushkin2012}.
The quantity accessible in a XPCS experiment is the intensity auto-correlation function $g_2(q,t)=\frac{\left\langle\left\langle I_{p}(q,t_1)I_{p}(q,t_1+t)\right\rangle_{p}\right\rangle}{\left\langle \left\langle I(q,t_1)\right\rangle_{p}\right\rangle\left\langle \left\langle I(q,t_1)\right\rangle_{p}\right\rangle}$, where $\left\langle...\right\rangle_{p}$ denotes the ensemble average over all the pixels of the detector and $\left\langle...\right\rangle$ is the temporal average. $g_2(q,t)$ can be related to the intermediate scattering function $f(q,t)$ through the Siegert relation $g_2(q,t)=1+B(q)\left|f(q,t)\right|^2$, with $q$ the scattering vector and $B(q)$ a setup-dependent parameter \cite{madsen2010}. By measuring the signal of all pixels, the intensity distribution is indeed sampled correctly allowing the extension of the Siegert relation to non-ergodic systems \cite{kirsch1996,bartsch1997,cipelletti1999}. We perform XPCS measurements at a wave-vector $q_0=2.56$ \r{A}$^{-1}$ corresponding to the maximum of the static structure factor, thereby probing directly the dynamics at the inter-particle distance $2\pi/q_0\sim2$ \r{A}. \\
Our sample was produced by melt spinning with an extremely fast quench ($10^6$ K/s) from the high-temperature melt down to $T=300$ K (see SI). We first measure this "as-quenched" glass (AQ-g) by performing several isothermal runs at increasing values of $T$, up to $T_g$. We then explore the supercooled liquid region ($T>T_g$), before slowly cooling again the system below $T_g$ ("slowly-cooled glass", SC-g). A second as-quenched glass (AQ-g2) is also measured, directly heated to 415 K without performing intermediate steps. Between isothermal runs $T$ was changed at a rate of 1 K/min. \\
Figure \ref{Fig1} shows a selection of normalized $g_2(t)$ measured for temperatures above and below the calorimetric $T_g$ for the SC-g. The data are reported together with the best fit obtained by using the Kohlrausch-Williams-Watt expression $g_2(t)=1+A\exp(-2(t/\tau)^{\beta})$, with $\tau$ the structural relaxation time, $\beta$ the shape parameter and $A=A(q,T)=B(q)*f_q(T)$, where $f_q(T)$ is the non-ergodicity factor appearing in the expression of the final decay of the intermediate scattering function, $f(q,t)=f_q(T)\exp(-t/\tau)^\beta$. A first important observation is that the dynamics is not frozen in the glassy state: $g_2(t)-1$ displays a full decay to zero, whose characteristic time $\tau$ increases with decreasing temperature. Thus, structural relaxation occurs even if the system has not fully equilibrated \cite{berthier2011}. A full decay is not completely reached at lower temperatures due to the large relaxation time of the sample. As discussed below, in the case of AQ-g the dynamics is much faster and we do observe a full decay at all $T$. More surprising, we find a remarkable equilibrium/out-of-equilibrium dynamical crossover at $T_g$ as signalled by a sharp change of the shape of $g_2(t)$, from stretched ($\beta=0.88<1$) in the supercooled liquid to "compressed" ($\beta\sim1.3>1$) in the glass. As it will be discussed later, the curves in the glass depend on waiting time and thermal history, which is not the case for data collected above 408 K.\\
Stretched exponential functions are a universal feature of supercooled liquids \cite{angell2000b}, reflecting a wide distribution of relaxation times, generally attributed to the existence of dynamical heterogeneities \cite{richert2002,berthier2011b}. Differently, the compressed behaviour found in our structural glass is totally unexpected and cannot be explained as resulting from a distribution of single exponential relaxation times, as it would be within the heterogeneous dynamics scenario. Indeed, this scenario could explain a growth of $\beta$ in the out-of-equilibrium state, but only up to $\beta\leq1$ \cite{richert2010,moyhinan1993,narayanaswamy1971,moynihan1976,lubchenko2004}. The idea would be that below $T_g$ only the regions characterized by a faster-than-average dynamics contribute to the decay of the correlation function, leading to a narrower distribution of relaxation times and thus to a $\beta$ closer to, but still smaller than, $\beta=1$, the limiting case of a single exponential decay. An alternative explanation for the growth of $\beta$ might be that the relaxation time evolves during the measurement of $g_2(t)$, altering the shape. We rule out this hypothesis by verifying that $\beta$ does not depend on the time interval over which $g_2(t)$ is averaged, and by noting that the slowing down of the dynamics below $T_g$ would lead to decrease the measured value of $\beta$, rather than increasing it \cite{lunkenheimer2005}. \\
Compressed correlation functions have been observed in out-of-equilibrium soft materials, such as concentrated colloidal suspensions and polymeric gels, and for nanoparticle probes in glass former matrix \cite{cipelletti2000,czakkel2011,bandyopadhyay2004,cipelletti2003,bouchaud2001,ballesta2008,caronna2008,guo2009}. In the latter case a similar stretched/compressed crossover was reported; however, those experiments probed length scales much larger than that of the glass former units and located the crossover in the liquid phase, well above $T_g$ \cite{caronna2008,guo2009}. By contrast, our experiments provide strong evidence of compressed exponential relaxations as soon as the system falls out of equilibrium below $T_g$. An increase of $\beta$ for $T<T_g$ was also reported in the investigation of the dielectric spectra of some polymeric glasses \cite{alegria1995,alegria1997}. However, in these systems $\beta$ increases by not more than $\sim$10$\%$ with respect to the corresponding liquid value, and always remains less than one. In our case, not only does $\beta$ increase by more than $\sim$40$\%$ in the SC-g (Figure \ref{Fig1}b), but it is also nearly temperature independent, in contrast with the prediction of a recent theoretical work \cite{lubchenko2004}.\\
Since diffusive processes driven by thermal energy lead to an exponential (or stretched exponential) structural relaxation, the decay observed here points to a different origin of the microscopic dynamics, unaccounted for in previous studies \cite{richert2010,moyhinan1993,narayanaswamy1971,moynihan1976,lubchenko2004}. We propose internal stress relaxation as the main physical mechanism responsible for the dynamics in the glass. Indeed, in experiments on soft materials a similar compressed exponential relaxation has been associated with ballistic rather than diffusive motion due to internal stresses stored in the system \cite{cipelletti2003,bouchaud2001}. These stresses can be described by progressive irreversible rearrangements that act as local sources of dipolar stress leading to a compressed exponential shape of the dynamic structure factor \cite{bouchaud2001}. It is well-known that the fast quenching used to produce metallic glasses leads to internal stresses or excess frozen in free volume \cite{busch1998,taub1980,beukel1990,evenson2011,yavari2005}, which may be released by heating the system (see Dynamic Mechanical Analyzer (DMA) and Differential Scanning Calorimetry (DSC) data in SI). \\
The temperature dependence of the structural relaxation times is shown in Figure \ref{Fig2}. Data corresponding to three different thermal paths are presented (full symbols): i) diamonds are taken from the curves reported in Figure \ref{Fig1} corresponding to the supercooled liquid and to the SC-g; ii) squares correspond to the AQ-g; iii) the star corresponds to AQ-g2. Macroscopic measurements obtained by viscosimetry, DMA and DSC are also shown as empty symbols. The viscosity data \cite{busch1998b} have been rescaled to the relaxation time values measured at higher temperatures with DMA. Data calculated from DSC measurements are also reported (see SI). 
The XPCS data taken above 403 K follow the macroscopic equilibrium liquid line and correspond to the stretched exponential correlation curves reported in Figure \ref{Fig1}, typical of supercooled liquids. All the points below the equilibrium line are instead in the glassy state. They come from compressed exponential curves and display aging. In the glassy state $\tau$ shows a strong cooling-rate-dependent departure from the behaviour observed at equilibrium, being the dynamics much faster in the AQ-g than in the SC-g.\\
On increasing $T$ in the AQ-g, the dynamics gets smoothly faster with $T$, until $\sim393$ K where $\tau$ increases sharply with increasing temperature, before decreasing again above 408 K, once the equilibrium curve is reached. This surprising increase is likely just a consequence of annealing during isothermal runs. Here the system slowly relaxes its internal stresses during heating and loses memory of its original thermal treatment, thus falling on the upper equilibrium curve \cite{hodge1994,bednarcik2011}. By contrast, if the system is continuously heated up to a higher temperature without having the time to rearrange (AQ-g2, in Figure \ref{Fig2}), this feature is not observed, and the glass meets its corresponding liquid at higher temperatures.\\
When the system falls out of equilibrium it starts aging. The age dependence of the dynamics is directly captured by XPCS through the analysis of the two-time correlation function $C$($t_1$,$t_2$) (Figure \ref{Fig3}). This quantity corresponds to the instantaneous correlation between two configurations at times $t_1$ and $t_2$. Lines parallel to the diagonal from the lower left to the upper right corners are iso-lag time lines ($t=t_1$-$t_2$=constant), while the sample age varies along such lines according to $t_{age}$ = ($t_1$+$t_2$)/2 \cite{madsen2010}. The slowing down of the dynamics is illustrated by the broadening of the yellow-red ridge along t=0 as the age increases. Since the evolution of the dynamics due to aging is slow, we time-average $C$($t_1$,$t_2$) over short time windows to obtain age dependent correlation functions $g_2(t,t_w)$. Here $t_w=t_0$+($t_f$-$t_i$)/2, $t_0$ being the delay of the start of the measurement from temperature equilibration, and $t_f$ and $t_i$ the final and starting times of the interval considered for averaging $C$($t_1$,$t_2$). Examples of different $g_2$($t$,$t_w$) measured for different ages are reported in Figure \ref{Fig3}b. On increasing $t_w$, the dynamics shifts towards longer time scales.\\
Figure \ref{Fig4}a reports the $t_w$ dependence of $\beta$ in the AQ-g for $T\geq$318 K. We find a temperature- and $t_w$-independent value $\beta$=1.5, which is $\sim$70$\%$ larger than in the supercooled liquid. This value is also larger than that for the SC-g ($\beta\sim1.3$), confirming that deviations from diffusive behaviour are enhanced by a faster quench, consistent with the internal stress picture. The value $\beta$=1.5 has been often observed in studies of soft materials \cite{cipelletti2000,cipelletti2003,bandyopadhyay2004} and it is also in agreement with the proposed models \cite{bouchaud2001,cipelletti2003}. Only at 403 K does $\beta$ decrease at very large $t_w$, presumably indicating that the sample gets closer to equilibrium.\\
Surprisingly, the aging differs from the linear or sublinear laws reported in the literature and exhibits remarkable scaling properties. For each temperature $\tau$ increases with $t_w$ exponentially, and the growth rate does not depend on $T$. This suggest that sets of data taken at the same temperature may be modelled by the empirical law\\ 
\begin{equation}
\label{eq1}
\tau(T,t_w)=\tau_0(T)exp(t_w/\tau^{\ast})
\end{equation}
and rescaled onto a master curve by plotting $\tau/\tau_0$ vs. $t_w$, as shown in Figure \ref{Fig4}b, where $t^{\ast}$=7940$\pm$70 s independent of $T$. An exponential growth of $\tau$ with sample age has been previously reported in a few soft materials \cite{cipelletti2000,schosseler2006,robert2006,masri2010}: our results show that this unexpected behavior may be more general than previously thought. The constant shape parameter and the perfect scaling of $\tau(T,t_w)/\tau_0(T)$ lead moreover to the validity of a time-waiting time-temperature superposition (TTTS) principle in the AQ-g and therefore to the possibility of obtaining a single master curve plotting $g_2(t)$ as a function of the reduced time $t/(\tau(T,t_w)/\tau_0(T))$.\\
The exponential aging persists even close to $T_g$, when $\beta$ starts to decrease with increasing $t_w$. Here the system enters an intermediate aging regime where both $\tau$ and $\beta$ vary with age. This is the case for the AQ-g data taken at 403 K (Figure \ref{Fig4}a) and for all data taken in the SC-g (see SI). Generally, one expects the exponential aging to stop at one point when the materials is partially annealed. To test this hypothesis, we lower the temperature of the AQ-g2 glass to 403 K, after partially annealing it at 415 K without equilibration (Figure \ref{Fig2}). At 403 K, the AQ-g2 data display the same fast aging as for the AQ-g, although with a higher starting value $\tau_{0,AQ-g2}$(T)=$390\pm20$ s and a slightly different $t^{\ast}$$_{AQ-g2}$=5600$\pm$200 s, due to the different thermal history. Differently from the AQ-g, upon increasing $t_w$ the AQ-g2 glass enters a second dynamical regime, where aging slows down so much that it becomes undetectable within our experimental window. Here $\tau$ is most likely reaching equilibrium sublinearly, as in the case of polymeric glasses and colloids \cite{struik1978,alegria1997,cipelletti2005}. Note that here $\beta\sim1.15$ is $t_w$ independent and still $\sim30\%$ higher than in the corresponding liquid, confirming that the sample is not fully equilibrated. \\
In conclusion, we have scrutinized the dynamics of a metallic glass on the atomic length scale and for different cooling rates and thermal paths. We find a universal dynamical crossover between the supercooled liquid phase and the glassy state where the shape of the correlation function changes sharply when crossing $T_g$. We have proposed that internal stress relaxation is the mechanism driving the dynamics in the glass. Consistently, we find that deviations from the diffusion-driven dynamics typically observed in the supercooled state are more important for samples where internal stresses are likely to be larger, e.g. samples submitted to a faster quench and at earlier ages. As the sample ages, internal stress are presumably relaxed, but in a way that depends on thermal history, leading to distinct aging regimes.\\
Different aging regimes and compressed correlation functions have been reported in the past for gels and colloids, thus suggesting a universal phenomenology \cite{cipelletti2000,czakkel2011,schosseler2006,robert2006,masri2010}.\\

We gratefully thank S. Capaccioli, W. Kob, A. Madsen, A. Robert, B. Capone and G. Baldi for stimulating discussions. 
We acknowledge H. Vitoux, K. L'Hoste, L. Claustre for the technical support for the XPCS experiments, G. Garbarino for the XRD measurements, and M. Reynolds for a careful proof reading of the manuscript. EP and PB acknowledge financial support from CICYT Grant No. MAT2010-14907 and Generalitat de Catalunya Grants No. 2009SGR1225 and No. 2009SGR1251.

\end{document}